\documentclass[%
 reprint,
nofootinbib,
 amsmath,amssymb,
 aps,
]{revtex4-1}

\usepackage{graphicx}
\usepackage{dcolumn}
\usepackage{bm}

\begin{document}

\preprint{}

\title{Gravity with free initial conditions: a solution to the
  cosmological constant problem testable by CMB B-mode polarization}

\author{Tomonori Totani}
\affiliation{%
Department of Astronomy, 
School of Science, The University of Tokyo \\
7-3-1 Hongo, Bunkyo-ku, Tokyo 113-0033, Japan
}%
\affiliation{
Research Center for the Early Universe,
School of Science, The University of Tokyo \\
7-3-1 Hongo, Bunkyo-ku, Tokyo 113-0033, Japan}%


\date{\today}

\begin{abstract}
In standard general relativity the universe cannot be started with
arbitrary initial conditions, because four of the ten components of
the Einstein's field equations (EFE) are constraints on initial
conditions.  In the previous work it was proposed to extend the
gravity theory to allow free initial conditions, with a motivation to
solve the cosmological constant problem. This was done by setting four
constraints on metric variations in the action principle, which is
reasonable because the gravity's physical degrees of freedom are at
most six.  However, there are two problems about this theory; the
three constraints in addition to the unimodular condition were
introduced without clear physical meanings, and the flat Minkowski
spacetime is unstable against perturbations. Here a new set of
gravitational field equations is derived by replacing the three
constraints with new ones requiring that geodesic paths remain
geodesic against metric variations.  The instability problem is then
naturally solved.  Implications for the cosmological constant
$\Lambda$ are unchanged; the theory converges into EFE with nonzero
$\Lambda$ by inflation, but $\Lambda$ varies on scales much larger
than the present Hubble horizon. Then galaxies are formed only in
small $\Lambda$ regions, and the cosmological constant problem is
solved by the anthropic argument.  Because of the increased degrees of
freedom in metric dynamics, the theory predicts new non-oscillatory
modes of metric anisotropy generated by quantum fluctuation during
inflation, and CMB B-mode polarization would be observed differently
from the standard predictions by general relativity.
\end{abstract}

\maketitle


\section{Introduction}
\label{sec:intro}

It is a well known fact that not all of the ten components of the
Einstein's field equations (EFE) are dynamical equations.  Because of
the freedom of general coordinate transformation, time evolution of
all the ten metric components $g_{\mu\nu}$ cannot be determined even
if $g_{\mu\nu}$ and their time derivatives are given at an initial
spacelike hypersurface\footnote{As usual, Greek indices run from 0 to
  3 and Latin indices are spatial from 1 to 3, with the sign
  convention same as ref. \cite{bakoten}. A partial derivative is
  denoted as $\partial_\mu$ or a comma, and a covariant derivative as
  $\nabla_\mu$ or a semicolon. The fundamental constants $c$ and
  $\hbar$ are set equal to unity.}.  The $0 \mu$ components of EFE do
not include second time derivatives of any metric component, and
$\partial_0^2 g_{0\mu}$ do not appear at all in EFE.  Therefore the
$0\mu$ components of EFE just give four constraints on initial
conditions, and the contracted Bianchi identities guarantee that they
hold at all time if they are satisfied at an initial spacelike
hypersurface.  Therefore, when a universe (spacetime) is born, only a
special set of physical states are allowed among all that are
otherwise physically possible.  For example, when we consider the
Friedmann-Lema\^itre-Robertson-Walker (FLRW) cosmology, the evolution
of the universe is simply determined by one of the four constraints
(i.e., the Friedmann equation), which does not include second time
derivative $\ddot a$ of the scale factor.  The Hubble parameter $H =
\dot a/a$ is determined once energy density and curvature are given,
though we can imagine a universe with various values of $H$.

From the viewpoint of the action principle to derive EFE, the
existence of such four constraints originates from considering
variations of all the ten metric components independently, while the
physical degrees of freedom (DOFs) of gravity are at most six because
of the four from coordinate transformation.  However, the
Einstein-Hilbert action is already invariant under coordinate
transformation, and we do not have to request a stationary action
condition about metric variations related to coordinate
transformation\footnote{An infinitesimal coordinate transformation of
  $x^\mu \rightarrow x^\mu + \zeta^\mu$ generates metric variations of
  $\delta g_{\mu\nu} = - \zeta_{\mu ; \nu} - \zeta_{\nu ; \mu}$, and the
  stationary action condition about this leads to the contracted Bianchi
  identities and the energy-momentum conservation law by a partial
  integration, rather than EFE.}.  This implies that there is a sort
of redundancy in deriving the full set of 10-component EFE, and it is
not unreasonable to expect that nature may disfavor such a
redundancy. Rather, it may be natural to derive gravitational field
equations by metric variations constrained into six physical DOFs of
gravity.  Then the four constraint equations would disappear, and such
a theory would be able to describe time evolution of spacetime
starting from any physically possible initial states.

Such a theory of gravity was proposed in the previous study
\cite{Totani:2016} (hereafter Paper I), adopting a guiding principle
that the gravity theory should be able to describe time evolution
starting from free initial conditions.\footnote{One may consider that
  a similar argument may also apply for electromagnetic dynamics,
  because there are similarities between mathematical structures of
  EFE and the Maxwell equations.  However, there is considerable
  difference in the nature of gravity and electromagnetic forces as
  well, and here we simply assume that this principle is valid only
  for gravity.}  The main motivation of this work was the cosmological
constant ($\Lambda$) problem (see e.g.
\cite{Frieman:2008,Caldwell:2009,Martin:2012,Weinberg:2012} for
reviews).  The key is how to set four constraints on metric variations
$\delta g_{\mu\nu}$ to extract six gravity DOFs. The unimodular
condition, $\delta (\sqrt{-g}) = (1/2) \sqrt{-g} \, g^{\mu\nu} \delta
g_{\mu\nu} = 0$, was chosen as the first one.\footnote{Note that we
  request only the variational condition of $\delta (\sqrt{-g}) = 0$,
  though the original unimodular condition was $\sqrt{-g} = 1$. This
  variational condition is sufficient to generate the cosmological
  constant term in aribtrary coordinate systems with $\sqrt{-g} \neq
  1$.}  This condition has been studied for a long time starting from
Einstein
\cite{Einstein:1919a,Anderson:1971,vanderBij:1981,Buchmuller:1988,Unruh:1989,Henneaux:1989,Ng:1990,Ng:1991,Finkelstein:2000,Ng:2001,Smolin:2009,Ellis:2011,Ellis:2014},
and it is interesting concerning the $\Lambda$ problem because a term
$\Lambda_g(x) \, g_{\mu\nu}$ appears in the field equations as a
Lagrange multiplier.  However, introducing only this constraint does
not solve the $\Lambda$ problem, because the contracted Bianchi
identities require $\Lambda_g$ to be a universal integration constant,
and we do not know how to set its initial value. In Paper I, the three
more constraints were assumed to be $\delta g_{0i} = 0$, simply to
make the $0i$ components of EFE ineffective.  This violates general
covariance, and for a consistent theory it was assumed that any
spacetime is initially created as a spacelike hypersurface, and the
conditions $\delta g_{0i} = 0$ hold in preferred reference frames,
i.e., synchronous coordinate systems starting from this hypersurface.

The motivation of just allowing free initial conditions may not be
strong enough to consider an alternative theory of gravity.  However,
the proposed theory gives a solution both for the smallness and
coincidence problems of $\Lambda$.  A universe is assumed to start
with highly inhomogeneous conditions, and hence $\Lambda_g(x)$ is not
a universal constant.  However, once a portion of the universe starts
inflation
\cite{Starobinsky_79,Kazanas_80,Guth_81,Sato_81,Linde_81,Albrecht+82},
anisotropy rapidly disappears and $\Lambda_g(x)$ converges into a
cosmological constant term $\Lambda$.  Hence the universe can be
described by EFE$+\Lambda$ within the present-day Hubble horizon, but
the final total value of $\Lambda \equiv \Lambda_g + \Lambda_{\rm
  vac}$ can be positive or negative and changes on scales much larger
than the Hubble horizon, where $\Lambda_{\rm vac}$ is the contribution
from vacuum energy density of all relevant fields in the particle
physics theory. Then the anthropic argument for the cosmological
constant \cite{Barrow:1986,Weinberg:1987} applies; galaxy formation is
possible only in the regions of $|\rho_\Lambda| \lesssim \rho_M$, like
concentration of human populations to coastal areas on Earth regarding
$\Lambda$ as altitude, where $\rho_\Lambda$ and $\rho_M$ are energy
densities of $\Lambda$ and matter, respectively. The probability
distribution of $\Lambda$ is expected to be flat per unit $\Lambda$
because $|\Lambda| \ll |\Lambda_g|$ and $|\Lambda_{\rm vac}|$.
Combined with galaxy formation efficiency as a function of $\Lambda$,
the probability of finding a small $\Lambda$ value as observed
\cite{Efstathiou+90,Fukugita+90,Yoshii_93,Krauss+95,Ostriker+95,Riess+98,Perlmutter+99,Spergel+03}
is not extremely small (typically $\sim$ 5--10 \% \cite{Sudoh:2016}).
Moreover, since the theory becomes EFE+$\Lambda$ in the present Hubble
horizon, it passes all the classical tests supporting general
relativity.

However, there are two problems in the theory proposed by Paper I.
One is that the three constraints of $\delta g_{0i} = 0$ were
introduced in a rather ad hoc way without clear physical background.
The other is that, as will be shown in \S \ref{sec:stability}, the
flat Minkowski spacetime becomes unstable against perturbations. The
purpose of this work is to present a new version of the gravitational
field equations (\S \ref{sec:new-approach}), which are similar but
modified from those in Paper I, by replacing the three constraints
with those having a more solid physical basis: geodesic paths are kept
geodesic against metric variations.  It will be found that the
instability problem naturally disappears in this new version.  Then
implications for cosmology will be discussed in \S
\ref{sec:cosmology}, in particular primordial metric fluctuations
produced by quantum effect during inflation.  Since this theory has
more DOFs of metric dynamics than standard general
relativity, new modes of metric anisotropy would be generated, which
may be tested by future observations such as B-mode polarization of
the cosmic microwave background radiation (CMB).

\section{Stability of the field equations in Paper I}
\label{sec:stability}

The gravitational field equations proposed in Paper I are
\begin{eqnarray}
G^{\mu\nu} - \kappa \, T^{\mu\nu} = \Lambda_g(x) \, g^{\mu\nu}
 + \Xi^{\mu\nu} \ ,  
\label{eq:PaperI}
\end{eqnarray}
where $G^{\mu\nu} = R^{\mu\nu} - (1/2) \, R \,g^{\mu\nu}$ is the
Einstein tensor, $\kappa = 8\pi G$, $G$ the Newton's gravitational
constant, and $T^{\mu\nu}$ the energy-momentum tensor of matter.  The
term including a scalar field $\Lambda_g(x)$ comes from the unimodular
condition on $\delta g_{\mu\nu}$, and the tensor $\Xi^{\mu\nu}$ comes
from the three conditions of $\delta g_{0i} = 0$.  In this theory, it
is assumed that any spacetime realized in nature is finite toward past
in timelike directions, and the initial spacelike hypersurface at the
birth of the universe can be defined at least as a classical theory.
Then we can define a synchronous coordinate system starting from this
hypersurface, in which $g_{00} = 1$ and $g_{0i} = 0$ throughout the
spacetime.  In this reference frame, it is assumed that $\delta
g_{0i}$ do not represent gravity DOFs, and hence the $0i$ components
of EFE simply become ineffective by $\Xi^{\mu\nu}$, which takes the
form of $\Xi^{00} = \Xi^{ij} = 0$ with three nonzero components of
$\Xi^{0i} = \Xi^{i0}$.  Since the initial hypersurface of constant
time is fixed, the synchronous condition allows only transformations
within spatial coordinates [$x'^i = f^i(x^j)$], and the form of
$\Xi^{\mu\nu}$ is unchanged by these. The term $\Xi^{\mu\nu}$ can be
converted into any coordinate systems if we define the ordinary tensor
transformation law, and then eqs. (\ref{eq:PaperI}) become covariant.
The theory is derived from the same Lagrangians as general relativity
and hence the stationary action condition does not depend on choice of
reference frames. However, the general principle of relativity is
violated because the constraints on $\delta g_{\mu\nu}$ to extract the
gravity DOFs have a preferred frame. Since the preferred frame can
physically be specified by the initial spacelike hypersurface, this is
a consistent theory to determine evolution of the gravitational fields
and spacetime.

However, there is a problem in these equations, which becomes apparent
when we consider linear perturbations from the flat Minkowski
spacetime.  In the synchronous frame, we consider perturbations of
$h_{ij}$, $\Lambda_g(x)$, and $\Xi^{\mu\nu}$, where $g_{\mu\nu} =
\eta_{\mu\nu} + h_{\mu\nu}$ and $\eta_{\mu\nu}$ is the metric of the
Minkowski spacetime as usual.  The contracted Bianchi identities of
eqs. (\ref{eq:PaperI}) lead to\footnote{The evolution of matter is
  determined by the least action principle for the matter action in a
  fixed spacetime geometry, and hence the matter equations of motion
  are not changed from the standard theory and the energy conservation
  $\nabla_\mu T^{\mu\nu} = 0$ holds.}  $\partial_\mu \Lambda_g \,
g^{\mu\nu} + \nabla_\mu \Xi^{\mu\nu} = 0$, and the $\nu=0$ and $i$
components are:
\begin{eqnarray}
\partial_0 \Lambda_g + \partial_i \Xi^{0i} &=& 0 \\
- \partial_i \Lambda_g + \partial_0 \Xi^{0i} &=& 0 \ .
\end{eqnarray}
Therefore we find
\begin{eqnarray}
\partial_0^2 \Lambda_g + \Delta \Lambda_g = 0 \ ,
\end{eqnarray}
where $\Delta$ is the three-dimensional Laplacian.  This is clearly
unstable, because a mode of wavenumber $k$ will grow
exponentially.  The field $\Lambda_g(x)$ is a scalar and hence this
result does not depend on choice of coordinate systems.  
When we consider a plane
wave propagating into the $x^3$ direction, 
the mode $(h_{11} + h_{22})$ is prohibited in the ordinary EFE because 
the 00 and 33 components give
\begin{eqnarray}
(h_{11} + h_{22})_{,33} = (h_{11} + h_{22})_{,00} = 0 \ .
\end{eqnarray}
However, in this theory because of the freedom about $\Lambda_g(x)$
we find 
\begin{eqnarray}
(h_{11} + h_{22})_{,00} + (h_{11} + h_{22})_{,33} = 0 \ ,
\end{eqnarray}
which also grows exponentially.  Therefore, even if the theory
predicts that inflation produces a spatially flat universe obeying
EFE+$\Lambda$ with negligible $\Xi^{\mu\nu}$ (Paper I), a small
perturbation will rapidly grow after inflation and the Minkowski
spacetime is not stable.  This is the problem that should be solved in
the new field equations presented below.

\section{A new approach to derive the
gravitational field equations}
\label{sec:new-approach}

\subsection{Variations keeping geodesics}

The constraints of $\delta g_{0i} = 0$ in Paper I were introduced in a
rather ad hoc way simply to make the $0i$ components of EFE
ineffective, and this may be the cause of the difficulty described in
\S \ref{sec:stability}. In this work the unimodular condition
is kept as one of the four constraints 
because it has a connection to the cosmological constant term, and see
also Paper I for some theoretical motivations to introduce this
condition in the action principle. Then we should seek for new three
constraints on $\delta g_{\mu\nu}$ on a more solid physical
basis. Here geodesics may be relevant, because the effective number of
components of the geodesic equations is indeed three. The principle
of equivalence tells us that gravity can be erased in any local
inertial frames, and hence it is reasonable to expect that a geodesic
should remain geodesic against metric variations related to the
physical DOFs of gravity. It is difficult to find covariant
constraints on $\delta g_{\mu\nu}$ keeping any geodesics, and here the
same assumption as Paper I is adopted that any spacetime realized in
nature starts from a well-defined spacelike hypersurface. The paths of
fixed spatial coordinates in the synchronous reference frame starting from
the initial hypersurface are geodesics, and they are natural
``backbone'' of spacetime when we consider evolution of spacetime to
the timelike direction.  Therefore it is assumed that metric
variations should keep these geodesics in the action principle to
derive the gravitational field equations.

Let $u^\mu = dx^\mu/d\tau$ be the four-velocity field of these
backbone geodesics $x^\mu(\tau)$, where $\tau$ is the proper time
along the geodesics.  Therefore $u^\alpha \nabla_\alpha u^\mu = 0$,
and in the synchronous frame $\tau = x^0$ and $u^\mu = (1, 0, 0, 0)$.
The condition of keeping geodesics is given by $\delta (u^\alpha
\nabla_\alpha u^\mu) = 0$ against $\delta g_{\mu\nu}$. Here, it should
be noted that the backbone paths $x^i(x^0)$ are kept unchanged by
$\delta g_{\mu\nu}$, and accordingly $u^\mu$ changes by $\delta
g_{\mu\nu}$ because a four-velocity includes proper time that depends
on metric, as $d\tau^2 = g_{\mu\nu} dx^\mu dx^\nu$.  Metric and four
velocity fields after variations are denoted as $\tilde g_{\mu\nu}
\equiv g_{\mu\nu} + \delta g_{\mu\nu}$ and $\tilde u^\mu \equiv u^\mu
+ \delta u^\mu$, respectively, and then variations of $u^\mu$ become
\begin{eqnarray}
\delta u^\mu &=& \delta\left( \frac{dx^\mu}{d\tau} \right)
= - \frac{\delta (d\tau)}{d\tau} u^\mu \\
&=&  - \frac{1}{2} \delta g_{\rho\sigma}u^\rho u^\sigma u^\mu
  \propto u^\mu \ .
\label{eq:delta-u}
\end{eqnarray}
Because $\tilde u_\mu \tilde u^\mu = 1$ holds after variation, we find
\begin{eqnarray}
\tilde u^\alpha \tilde \nabla_\alpha (\tilde u_\mu \tilde u^\mu) 
= 2 \tilde u_\mu \tilde u^\alpha \tilde \nabla_\alpha \tilde u^\mu = 0 \ ,
\end{eqnarray}
where $\tilde \nabla_\alpha$ is a covariant derivative using
$\tilde g_{\mu\nu}$. Note that $\tilde u^\alpha \tilde \nabla_\alpha
\tilde u^\mu$ is not necessarily zero, if the path is no longer
geodesic after variation.  Therefore,
\begin{eqnarray}
\tilde u_\mu \tilde u^\alpha \tilde \nabla_\alpha \tilde u^\mu 
 = \delta( u_\mu u^\alpha \nabla_\alpha u^\mu ) 
 =  u_\mu \delta( u^\alpha \nabla_\alpha u^\mu ) 
 = 0 \ ,
\label{eq:orthogonal}
\end{eqnarray}
i.e., variations of the geodesic equations are perpendicular to
$u^\mu$. This means that the condition of keeping geodesics consists of three
constraints rather than four, which is not surprising because there
are only three independent components in the geodesic equations.

Variations of the geodesic equations can be written with
$\delta u^\mu$ and $\delta \Gamma^\mu_{\rho\sigma}$
as
\begin{eqnarray}
\delta( u^\alpha \nabla_\alpha u^\mu ) &=& 
\delta u^\alpha \ \nabla_\alpha u^\mu 
+  u^\alpha \nabla_\alpha (\delta u^\mu) \nonumber \\
&& + \ u^\alpha u^\beta \delta \Gamma^\mu_{\alpha\beta} \ ,
\label{eq:del-geo1}
\end{eqnarray}
where $\Gamma^\mu_{\rho\sigma}$ are the Christoffel symbols.  Using
eq. (\ref{eq:delta-u}) and the geodesic equations, the first term on
the right hand side vanishes and we find
\begin{eqnarray}
\delta( u^\alpha \nabla_\alpha u^\mu ) = 
-\frac{1}{2} u^\alpha u^\rho u^\sigma u^\mu
  \nabla_\alpha \delta g_{\rho\sigma} 
+ u^\alpha u^\beta \delta \Gamma^\mu_{\alpha\beta} \ .
\label{eq:del-geo2}
\end{eqnarray}
Because this is perpendicular to $u^\mu$ and the first term on the
right hand side is proportional to $u^\mu$, in the synchronous frame
these constraints are equivalent to requesting that the spatial components
of the second term are zero, i.e.,
\begin{eqnarray}
2 u^\alpha u^\beta \delta \Gamma^i_{\alpha\beta} &=& g^{ij} 
(2 \partial_0 \delta g_{0j} - \partial_j \delta g_{00}) = 0 \ .
\end{eqnarray}
Therefore the three constraints can simply be expressed as
\begin{eqnarray}
2 \partial_0 \delta g_{0i} - \partial_i \delta g_{00} = 0 \ .
\label{eq:geodesic-constraint-sync}
\end{eqnarray}

\subsection{The new field equations}
\label{sec:new-field-eqs}

Here a new set of field equations is derived from the action principle
by adopting the constraints (eq. \ref{eq:geodesic-constraint-sync}) in
the synchronous coordinate system, instead of $\delta g_{0i} = 0$ in
Paper I.  An important feature of the constraints
(eq. \ref{eq:geodesic-constraint-sync}) is that they include
derivatives of metric, in contrast to another constraint of the
unimodular condition. In derivation of local differential field
equations generally, variation of the action $\delta S$ is expressed
with $\delta g_{\mu\nu}$ (but without their derivatives using partial
integrations), and then $\delta S = 0$ is requested for a local metric
variation that is nonzero only at an infinitesimal region. A
constraint without derivatives can easily be incorporated in such a
local metric variation, but it is not simple when a constraint
includes derivatives, leading to nonlocal field equations.

Variation of the action by $\delta g_{\mu\nu}$ in general relativity
is
\begin{eqnarray}
\delta S = \frac{1}{2 \kappa}
\int \left( G^{\mu\nu} - \kappa T^{\mu\nu} \right)
\sqrt{-g} \, \delta g_{\mu\nu} \, d^4x  \ ,
\label{eq:delta-S-GR}
\end{eqnarray}
and define $\xi^i(x)$ in a synchronous coordinate system
as integrations of the $0i$
components over time ($t \equiv x^0$), as
\begin{eqnarray}
\xi^i \sqrt{-g} &\equiv&
\int^t_{t_s} \left( G^{0i} - \kappa T^{0i} \right) \sqrt{-g} \ dt' 
+ C^i \ ,
\label{eq:def-xi-i}
\end{eqnarray}
where integration is over a path of constant spatial coordinates
(i.e., a backbone geodesic) starting from the initial hypersurface on
which time is $t_s$, and $C^i(x^j) = \xi^i(t_s, x^j) \sqrt{-g}$ are
integration constants.  Then the $0i$ and $i0$ metric variations in
eq. (\ref{eq:delta-S-GR}) can be converted into a variation about
$\delta g_{00}$ using the constraint
(\ref{eq:geodesic-constraint-sync}) as
\begin{eqnarray}
&2& \int \left(
  G^{0i} - \kappa T^{0i} \right) \sqrt{-g} \, \delta g_{0i} \, d^4x \\
&& \hspace{2cm} = 2\int \partial_0 (\xi^i \sqrt{-g}) \, \delta g_{0i} \, d^4x \\
&& \hspace{2cm} = - 2 \int \xi^i \, \sqrt{-g} \ \partial_0 \delta g_{0i} 
    \, d^4x \\
&& \hspace{2cm} = - \int \xi^i \, \sqrt{-g} \  \partial_i \delta g_{00} 
    \, d^4x \\
&& \hspace{2cm} = \int \frac{\partial_i (\xi^i \sqrt{-g})}{\sqrt{-g}}
    \, \sqrt{-g} \, \delta g_{00} \, d^4x \ ,
\end{eqnarray}
where the boundary terms are assumed to vanish as usual.
Now the stationary action condition about $\delta g_{00}$
and $\delta g_{ij}$, and the definition of $\xi^i$ (eq. \ref{eq:def-xi-i})
can be combined into ten-component new field equations as follows:
\begin{eqnarray}
G^{\mu\nu} - \kappa \, T^{\mu\nu} = \Lambda_g \, g^{\mu\nu}
+ \Xi^{\mu\nu}
\label{eq:new-field-eqs}
\end{eqnarray}
where
\begin{eqnarray}
\Xi^{00} &=& - \frac{\partial_i (\xi^i \sqrt{-g})}{\sqrt{-g}} 
 = - \frac{\partial_i \tilde \xi^i}{\sqrt{-g}} 
 = - \partial_i \xi^i - \Gamma^\alpha_{i\alpha} \, \xi^i \ ,
\label{eq:Xi_1} \\
\Xi^{0i} &=& \frac{\partial_0 (\xi^i \sqrt{-g})}{\sqrt{-g}} 
 = \frac{\partial_0 \tilde \xi^i}{\sqrt{-g}} 
 = \partial_0 \xi^i + \Gamma^\alpha_{0\alpha} \, \xi^i  \ ,
\label{eq:Xi_2} \\
\Xi^{ij} &=& 0 \ .
\label{eq:Xi_3}
\end{eqnarray}
Here, the unimodular condition is also adopted to generate the term
$\Lambda_g(x) \, g^{\mu\nu}$, and $\tilde \xi^i \equiv \xi^i \sqrt{-g}$.
Note that now $\Xi^{\mu\nu}$ is different from that in Paper I;
it is similar but $\Xi^{00}$ is nonzero and related to
$\Xi^{0i}$ through $\xi^i$ in this new version. 
Since $\xi^i$ is a time integration of the $0i$ component of EFE,
these are integro-differential equations about time, i.e., nonlocal.

If we define a tensor $\Xi^{\mu\nu}$ by the ordinary transformation
law, the field equations (\ref{eq:new-field-eqs}) would become a
covariant form holding in any coordinate systems.  However, the above
expression of $\Xi^{\mu\nu}$ by $\xi^i$
(eqs. \ref{eq:Xi_1}--\ref{eq:Xi_3}) is valid in synchronous coordinate
systems, and there is a freedom of coordinate transformation within
the spatial coordinates, $x'^i = f^i(x^j)$, even if the initial
spacelike hypersurface is fixed. Therefore if $\Xi^{\mu\nu}$ can be
defined as a tensor, eqs. (\ref{eq:Xi_1})--(\ref{eq:Xi_3}) should be
common in all these synchronous systems under an appropriate
transformation law of $\xi^i$, which can be shown as follows.  The
transformation law of $\Xi^{\mu\nu}$ by the spatial coordinate
transformation should be
\begin{eqnarray}
\Xi'^{00} &=& \Xi^{00} \label{eq:Xi_transf1}  \\
\Xi'^{0i} &=& \frac{\partial x'^i}{\partial x^j} \, \Xi^{0j} \\
\Xi'^{ij} &=& \frac{\partial x'^i}{\partial x^k} 
\frac{\partial x'^j}{\partial x^l}  \, \Xi^{kl} = 0 \ .
\end{eqnarray}
In a synchronous frame, the Christoffel symbol in eq. (\ref{eq:Xi_2})
is $\Gamma^\alpha_{0\alpha} = (1/2) \, g^{ij} g_{ij,0}$, which is
invariant under spatial coordinate transformations.  Then $\Xi^{0i}$
obeys the tensor transformation law if $\xi^i$ are transformed as
spatial components of a four-vector $\xi^\mu$ whose 0-th component is
$\xi'^0 = \xi^0 = 0$.  Using this four vector, eq. (\ref{eq:Xi_1})
becomes $\Xi^{00} = - \nabla_\mu \xi^\mu$, which is a scalar and hence
consistent with the transformation law of $\Xi^{00}$
(eq. \ref{eq:Xi_transf1}). Therefore
$\Xi^{\mu\nu}$ is consistently expressed by
eqs. (\ref{eq:Xi_1})--(\ref{eq:Xi_3}) in all the synchronous
coordinate systems starting from the initial hypersurface.  Finally,
if we define $\Xi^{\mu\nu}$ in non-synchronous coordinate systems as
those transformed from the synchronous systems by the ordinary tensor
law, a tensor $\Xi^{\mu\nu}$ is defined consistently in any coordinate
systems.

\subsection{Ambiguity in spacetime determination from
initial conditions and its removal}
\label{section:ambiguity}

The role of a gravitational theory should be to determine time
evolution of spacetime, i.e., $g_{ij}$ in a synchronous frame, when
$g_{ij}$, $\partial_0 g_{ij}$, and $T^{\mu\nu}$ on the initial
hypersurface are given.  However, the equations
(\ref{eq:new-field-eqs}) are not yet sufficient.  The second time
derivatives $\partial_0^2 g_{ij}$ appear only in the $ij$ components,
but these also include $\Lambda_g(x)$. Though $\Lambda_g$ can be
erased using the 00 component, $\partial_i \tilde \xi^i$ then appears
in the $ij$ components, which must be determined independently.  One
may consider that the three constraints from the $0i$ components
determine $\partial_i \tilde \xi^i$, but the $0i$ components determine
only $\partial_0 \tilde \xi^i$, with no constraints on $\tilde \xi^i$.
If we consider $\xi^i$ as new fields of physical DOFs, such an
ambiguity may not be a problem. However, our motivation is just to
remove the constraints on initial conditions in EFE with increased
DOFs of metric dynamics, rather than to complicate the gravity theory
by introducing new physical fields.  If there is another constraint on
the initial conditions of $\Lambda_g$ and/or $\partial_i \tilde\xi^i$,
the spacetime evolution would be unambiguously determined without new
physical fields.

This ambiguity arises because the field equations include both
$\tilde\xi^i$ and their time derivatives.  This is a consequence of
the nonlocal nature of the constraints to keep geodesics, which lead
to the field equations including integrations of the $0i$ components
of EFE over the past ($\tilde \xi^i$).  The nonlocality is somewhat
concordant with the basic assumption of the theory that there are
preferred reference frames defined by the initial spacelike
hypersurface, which is also a global property of the spacetime.  Then
it would be natural to relate the initial conditions about $\tilde
\xi^i$ and $\Lambda_g$ to the initial spacelike hypersurface. A
natural initial condition is that dynamical equations about
$\partial_0^2 g_{ij}$ are not affected by $\tilde \xi^i$ or
$\Lambda_g$ on the initial hypersurface.  This is achieved by
requesting a covariant condition $\Lambda_g = 0$ on the initial
hypersurface, because the effect of $\tilde \xi^i$ is propagated to
the $ij$ components through $\Lambda_g$ in the 00 component.\footnote{
  Another possible constraint to remove the ambiguity is setting
  $\Xi^{00} = -\nabla_\mu \xi^\mu = 0$, rather than $\Lambda_g = 0$ on
  the initial hypersurface. However, in this option fluctuation of
  $\Lambda_g$ produced by quantum effect during inflation would be too
  large to be consistent with the presently observed universe (see
  \S\ref{sec:b-mode}). }  Therefore, in this theory it is assumed that
any spacetime realized in nature is born as a spacelike hypersurface
on which $\Lambda_g = 0$, by physics beyond the level of a classical
theory.

Now the theory can determine spacetime evolution without ambiguity
from given initial conditions about $g_{ij}$, $\partial_0 g_{ij}$, and
$T^{\mu\nu}$.  The principle of free initial conditions is satisfied
because of the freedom about $\partial_i \tilde \xi^i$ and $\partial_0
\tilde \xi^i$ in the 00 and $0i$ components, respectively.  The
initial value of $\partial_i \tilde \xi^i$ is determined by the 00
component with $\Lambda_g = 0$, and evolution of $\tilde \xi^i$ is
determined by the $0i$ components.  The metric evolution is determined
by $\partial_0^2 g_{ij}$ of the $ij$ components, and then $\Lambda_g$
evolution is determined by the time derivative of the 00 component.
Therefore the $ij$ components to determine $\partial_0^2 g_{ij}$ are
the same as EFE on the initial hypersurface, and deviation from EFE
emerges by evolution of $\Lambda_g$ from zero, which is related to
nonlocal integration of physical quantities over the past backbone
paths starting from the initial hypersurface.  Note that $\xi^i$
affect the metric evolution only through the form of $\partial_i
\tilde \xi^i$, and hence any $\xi^i$ on the initial hypersurface give
the same spacetime solution if $\partial_i \tilde \xi^i$ is the same.

\subsection{Covariant derivation by Lagrange multipliers}

The fact that $\Xi^{\mu\nu}$ becomes a tensor using the four-vector
field $\xi^\mu$ implies that a more covariant derivation of the field
equations (\ref{eq:new-field-eqs}) in general coordinate systems may
be possible, and indeed it is, using the concept of Lagrange
multipliers as shown below. An important difference from the case of
the unimodular condition is that the constraints to keep geodesics
include metric derivatives, and hence derivatives of multipliers
appear in the field equations by partial integrations, which is
related to the nonlocal nature of the constraints as already
discussed.  If the gravity plus matter action $S$ is stationary
against $\delta g_{\mu\nu}$ under the constraints of keeping
geodesics, $\delta(u^\alpha \nabla_\alpha u^\mu) = 0$, a modified
action $\tilde S \equiv S + S_C$ should be stationary with a Lagrange
multiplier four-vector $\xi^\mu$, where
\begin{eqnarray}
S_C &=& \frac{1}{\kappa} 
\int \xi_\mu \, u^\alpha \nabla_\alpha u^\mu \sqrt{-g} \; d^4x 
\end{eqnarray}
and a factor of $1/\kappa$ is introduced to make the resultant field
equations equivalent to eqs. (\ref{eq:new-field-eqs}).  Because of the
orthogonality between $u^\mu$ and $\delta (u^\alpha \nabla_\alpha
u^\mu)$, a condition of $\xi_\mu u^\mu = 0$ can be set.  Variation
of $S_C$ by $\delta g_{\mu\nu}$ is then
\begin{eqnarray}
\delta S_C &=& \frac{1}{\kappa} 
\int \xi_\mu \, \delta (u^\alpha \nabla_\alpha u^\mu) \sqrt{-g} \, d^4x \\
&=& \frac{1}{\kappa} \int \xi_\mu u^\alpha u^\beta \delta 
 \Gamma^\mu_{\alpha\beta} \sqrt{-g} \ d^4x \ ,
\end{eqnarray}
where the contribution from the first term on the right hand side of
eq. (\ref{eq:del-geo2}) vanishes by $\xi_\mu u^\mu = 0$, and we do not
have to consider variation of $\sqrt{-g}$ because of the geodesic
equations for unvaried quantities.  Since variations of the Christoffel
symbols $\delta \Gamma^\mu_{\alpha\beta}$ are a tensor, 
$\delta S_C$ should be a scalar.  We find
\begin{eqnarray}
\delta S_C &=& \frac{1}{2\kappa} \int \Bigl\{ \xi_\mu u^\alpha u^\beta 
 \bigl[ \; g^{\mu\rho} ( \delta g_{\rho\beta,\alpha}
 + \delta g_{\alpha\rho,\beta} - \delta g_{\alpha\beta,\rho} ) \nonumber \\
&& \hspace{1cm} + \ \delta g^{\mu\rho} (g_{\rho\beta,\alpha} 
 + g_{\alpha\rho,\beta} - g_{\alpha\beta,\rho}) \; \bigr] \Bigr\}
 \ \sqrt{-g} \ d^4 x \nonumber \\
&=& - \frac{1}{2\kappa}
\int \Bigl[  \nabla_\alpha (\xi^\rho u^\alpha u^\beta) \delta g_{\rho\beta}
  + \nabla_\beta (\xi^\rho u^\alpha u^\beta) \delta g_{\alpha\rho} \nonumber \\
  && \hspace{1cm} - \nabla_\rho (\xi^\rho u^\alpha u^\beta) \delta g_{\alpha\beta} 
      \nonumber \\
  && \hspace{1cm} + \ {(\rm terms \ including \ \Gamma^\mu_{\rho\sigma})} 
  \ \Bigr] \sqrt{-g} \ d^4x \ ,
\end{eqnarray}
where partial integrations with vanishing boundaries have been used to
get the second line, and the terms including Christoffel symbols
appear by converting partial derivatives into covariant ones and
$\partial_\alpha(\sqrt{-g}) = \Gamma^\mu_{\mu\alpha} \sqrt{-g}$.
Since $\delta S_C$ is a scalar, the terms including Christoffel symbols
should vanish, which of course can be checked by a direct
calculation. Therefore,
\begin{eqnarray}
\delta S_C 
&=& - \frac{1}{2\kappa} \int \Xi^{\mu\nu} \delta g_{\mu\nu} \sqrt{-g} \ d^4x
\label{eq:geod_constr_integ2}
\end{eqnarray}
where
\begin{eqnarray}
\Xi^{\mu\nu} \equiv  \nabla_\rho (\xi^\mu u^\nu u^\rho) 
  + \nabla_\rho (\xi^\nu u^\mu u^\rho)
  - \nabla_\rho (\xi^\rho u^\mu u^\nu) \ .
\end{eqnarray}
Then the field equations become eq. (\ref{eq:new-field-eqs}) after
adding the unimodular condition.  It is straightforward to verify by
component calculations that this tensor $\Xi^{\mu\nu}$ is exactly the
same as eqs. (\ref{eq:Xi_1})--(\ref{eq:Xi_3}) in a synchronous
coordinate system.

\subsection{Linear perturbation analysis}
\label{sec:linear_theory}

To examine the stability of flat spacetime in this theory, consider
linear perturbations about $h_{ij}$, $\Lambda_g$ and $\xi^i$ from the
Minkowski spacetime without matter in a synchronous coordinate.
The contracted Bianchi identities
$\nabla_\mu (\Lambda_g \, g^{\mu\nu} + \Xi^{\mu\nu}) = 0$ for
the 0 and $i$ components are:
\begin{eqnarray}
\partial_0 \Lambda_g + \partial_0(-\partial_i \xi^i)
 + \partial_i(\partial_0 \xi^i) = \partial_0 \Lambda_g = 0 \ , \\
- \partial_i \Lambda_g + \partial_0^2 \xi^i = 0 \ .
\end{eqnarray}
Therefore at the level of linear perturbation, the scalar field
$\Lambda_g(x)$ is always a constant along the time coordinate. This is
in sharp contrast to the theory proposed in Paper I in which $\Lambda_g$
exponentially grows, and hence the instability problem
has disappeared. 

Next we consider plane wave perturbations along
the $x^3$ direction, and hence derivatives about $x^1$ and $x^2$
disappear.  The 10 components of the field equations 
(multiplied by a factor of 2) become
\begin{eqnarray}
00&:& \  (h_{11} + h_{22})_{,33} = 2 \, \Lambda_g  -  2 \, \xi^3_{,3} 
 \label{eq:linear_1} \\
11&:& \  \square h_{22} + h_{33,00} = - 2 \, \Lambda_g \\
22&:& \  \square h_{11} + h_{33,00} =  - 2 \, \Lambda_g  \\
33&:& \  (h_{11} + h_{22})_{,00} =  - 2 \, \Lambda_g  \\
12&:& \  - \square h_{12} = 0 \\
13&:& \  - h_{13,00} = 0 \\
23&:& \  - h_{23,00} = 0 \\
01&:& \  - h_{13,03} = - 2 \, \xi^1_{,0} \\
02&:& \  - h_{23,03} = - 2 \, \xi^2_{,0} \\
03&:& \  (h_{11} + h_{22})_{,03} = - 2 \, \xi^3_{,0} \ ,
  \label{eq:linear_10}
\end{eqnarray}
where $\square \equiv \eta^{\mu\nu} \partial_\mu \partial_\nu$ and
the superscripts in eqs. (\ref{eq:new-field-eqs}) have been
converted into subscripts by the ordinary tensor laws, and note that
$\Xi_{00} = \Xi^{00}$ and $\Xi_{0i} = g_{ij} \Xi^{0j} \sim - \Xi^{0i}$
up to the first order perturbations. It can be seen that the two modes
of gravitational waves ($h_{12}$ and $h_{11} - h_{22}$) in EFE are
unchanged.

Since free initial conditions are allowed in this theory, we consider
plane wave perturbations $h_{ij} = A_{ij} \exp(ikx^3)$ for all the
$ij$ components on the initial hypersurface. For simplicity, here
particular solutions with initial conditions of $\Lambda_g = 0$ and
$h_{ij, 0} = 0$ are considered\footnote{ Though this theory assumes
  $\Lambda_g = 0$ on the initial hypersurface when a spacetime was
  born, nonzero fluctuations of $\Lambda_g$ may be produced in later
  nonlinear cosmological evolution. However, as argued in \S
  \ref{sec:b-mode}, these conditions are expected to hold for
  fluctuations generated by inflation.}.  Then because $\partial_0
\Lambda_g = 0$, $\Lambda_g$ is zero throughout the spacetime, making
the $ij$ components the same as EFE.  The difference from EFE appears
as $\xi^i_{,0}$ and $\xi^3_{,3}$ only in the $0\mu$ components.  There
is no acceleration for $h_{13}$ and $h_{23}$, but the difference from
EFE is that their first time derivatives can be nonzero. However, we
have assumed $h_{ij, 0} = 0$ initially for the particular solutions
considered here, and hence they are kept constant with time. Then
these two modes are not different from those possible in EFE, and they
can be erased by a linear infinitesimal coordinate transformation
$x'^\mu = x^\mu + \zeta^\mu$ and $h'_{ij} = h_{ij} - \zeta_{i,j} -
\zeta_{j,i}$, where $\zeta_\mu \equiv \eta_{\mu\nu} \zeta^\nu$.  For
example, $h_{13}$ disappears by a spatial coordinate transformation
keeping the synchronous condition, $\zeta_1 = A_{13} (ik)^{-1}
\exp(ikx^3)$.

Now give a look at the mode $h_{11} + h_{22}$, which is also force
free and hence $A_{11} + A_{22}$ becomes constant by the assumed
initial condition of $h_{ij,0} = 0$.  It should be noted that this
mode is prohibited in EFE because of the 00 component, $(h_{11} +
h_{22})_{,33} = 0$, in contrast to $h_{13}$ and $h_{23}$.  But in this
theory $A_{11} + A_{22}$ can be nonzero by the freedom of
$\xi^3_{,3}$.  This mode cannot be erased by coordinate
transformations, and it may have observational consequences that are
different from EFE.  Finally, the mode $h_{33}$ feels a constant
acceleration as
\begin{eqnarray}
2 h_{33,00} &=& - 2 \xi^3_{,3} = (h_{11} + h_{22})_{,33}  \ .
\end{eqnarray}
Though it seems to evolve in time as $h_{33} = (Bt^2 + A_{33})
\exp(ikx^3)$ where $A_{33}$ is constant and $B = - (A_{11} + A_{22})
k^2/4$, this depends on the choice of a coordinate system. In fact,
this mode is transformed into
\begin{eqnarray}
h'_{00} &=& - 2 B \, \frac{1}{k^2} \, \exp(ikx^3) \\
h'_{03} &=& h'_{33} = 0
\end{eqnarray}
by a coordinate transformation of
\begin{eqnarray}
\zeta_0 &=& 2 B t \, \frac{1}{k^2} \, \exp(ikx^3) \\
\zeta_3 &=& (Bt^2 + A_{33}) \, \frac{1}{ik} \, \exp(ikx^3) \ .
\end{eqnarray}
This is no longer a synchronous coordinate system, but the mode
becomes a static gravitational field.  The initial fluctuation
$A_{33}$ disappears in $h'_{00}$, and $B$ is related to the mode
$(h_{11} + h_{22})$, and hence the mode $h_{33} $ is essentially the
same as $(h_{11} + h_{22})$.  This implies that the growing mode
$h_{33} \propto t^2$ is an artifact by the constrained choice of
synchronous coordinate systems, which becomes a static perturbed
Minkowski spacetime in non-synchronous frames.

\section{Implications for cosmology}
\label{sec:cosmology}

\subsection{Solving the cosmological constant problem}
\label{sec:Lambda-problem}

Implications for the cosmological constant problem are not much
changed from those obtained in Paper I. It is reasonable to assume
that the universe started from a highly inhomogeneous state without
the four constraints of EFE on initial conditions. In the theory
$\Lambda_g = 0$ is assumed on the initial hypersurface, but the linear
perturbation theory cannot be adopted in such a state and hence
$\Lambda_g(x)$ would soon evolve to nonzero values. Then in some
regions of the spacetime inflation begins by the vacuum energy density
of an inflaton field with a Hubble parameter value of $H$.  The
inflating regions become isotropic and homogeneous, and hence
anisotropic quantities such as $\xi^i$ should become unimportant, and
the contracted Bianchi identities ensure that $\Lambda_g(x)$ converges
to a universal constant. This should be examined by numerical studies
in future work, but can be verified at the linear theory level in the
flat FLRW metric as follows. Consider linear perturbations about
$h_{ij}$, $\delta\Lambda_g$, and $\xi^i$ from the FLRW metric in a
synchronous gauge, ignoring matter perturbations for simplicity.  Here
$h_{ij}$ are defined by $g_{ij} = -a^2 (\delta_{ij} - h_{ij})$ so that
they become the same as those in \S \ref{sec:stability} and \S
\ref{sec:linear_theory} when $a = 1$. The fluctuation $\delta
\Lambda_g(x)$ is discriminated from the nonzero constant $\Lambda_g$
of the background metric.  In this case
\begin{eqnarray}
\Xi^{00} &=& - \partial_i \xi^i \\
\Xi^{0i} &=& \partial_0 \xi^i + 3 H \xi^i
\end{eqnarray}
and the contracted Bianchi identities
$\nabla_\mu(\delta \Lambda_g \, g^{\mu\nu} + \Xi^{\mu\nu}) = 0$ become
\begin{eqnarray}
\nu = 0: \ \ \ &&
\partial_0 \delta \Lambda_g = 0 \\
\nu = i: \ \ \ &&
-\frac{1}{a^2} \partial_i \delta \Lambda_g  + \partial_0 \Xi^{0i}
+ 5 H \Xi^{0i} = 0   \ .
\end{eqnarray}

Therefore $\delta \Lambda_g$ becomes constant about time in the linear
theory also in the FLRW metric. From the second equation, $\Xi^{0i}$
should exponentially decay if $| 5H \Xi^{0i} | \gtrsim | \partial_i
\delta \Lambda_g / a^2 |$, and hence $| 5H \Xi^{0i} |$ is limited
below $| \partial_i \delta \Lambda_g / a^2 |$.  Consider a fluctuation
whose comoving wavenumber is $k_i \sim H$ at the beginning of
inflation, at which $a$ is normalized to unity.  Then we find
$|\Xi^{0i}| \lesssim |\delta \Lambda_g|/a^2$ and $|\Xi^{00}| \sim
|\Xi^{0i}|$, meaning that $\Xi^{\mu\nu}$ is unimportant compared with
$\delta \Lambda_g \, g^{\mu\nu}$ (i.e., $|\Xi^{00}| \ll |\delta
\Lambda_g|$ and $|\Xi^{0i}| \ll |\delta \Lambda_g|/a$) when $a \gg 1$.
Therefore fluctuations at the beginning of inflation rapidly disappear
and the field equations converge into EFE with a nonzero cosmological
constant $\Lambda$ on scales much smaller than the comoving scale of
initial inhomogeneity before inflation ($k_i^{-1}$). The symmetry of
general covariance is thus spontaneously restored. However, the
fluctuations of $\Lambda_g$ on the comoving scale $k_i^{-1}$ remain
after inflation, with the final value of $\Lambda = \Lambda_g +
\Lambda_{\rm vac}$ in the presently observed universe.  Then the
cosmological constant problem is solved by the anthropic argument, as
explained in \S \ref{sec:intro} (see Paper I for more discussions).

\subsection{Primordial metric anisotropy generated by inflation}
\label{sec:b-mode}

In \S \ref{sec:Lambda-problem} it was shown that the observable
universe should be described by EFE+$\Lambda$ after inflation at the
level of a classical theory. It is the standard paradigm that the
large scale structures observed in the present universe are generated
by quantum fluctuations of the inflaton field, and the two modes of
primordial gravitational waves ($h_{12}$ and $h_{11} - h_{22}$)
are also generated by quantum fluctuations of metric, which may be
observed as B-mode polarization of CMB (see \cite{Kamionkowski:2015}
for a review).  These quantum fluctuations are assumed to become
classical fluctuations when the wavelength of a mode becomes larger
than the Hubble horizon during inflation, though there are some
fundamental theoretical issues about such a process.
 
In the proposed theory, the four constraints on initial conditions in
EFE have been removed, and DOFs of metric dynamics are increased.  As
shown in \S \ref{sec:linear_theory}, the two gravitational wave modes
are not changed in the new theory.  The newly increased modes do not
obey the wave equation, but quantum fluctuations should exist for any
dynamical DOF. Therefore if the standard paradigm of quantum
generation of density and metric fluctuations is correct, all the six
modes of $h_{ij}$ fluctuations are expected to appear.  The amplitude
of primordial gravitational waves is predicted to be $\Delta_h^2 \sim
k^3 \langle |h_k|^2 \rangle \sim GH^2$, where $\langle |h_k|^2
\rangle$ is the power spectrum of metric fluctuations of a comoving
wavenumber $k$.  This can be derived by the uncertainty principle
about time ($H^{-1}$) and kinetic energy within the Hubble volume
($\dot h^2 / G \times H^{-3}$) using $\dot h \sim Hh$.  The same
argument can be applied to the new modes, and hence they are expected
to have similar amplitudes to those of the two gravitational wave
modes. An important difference is, however, that the new modes are not
oscillatory.  Matter fluctuations and gravitational waves are
predicted to be Gaussian, because the ground state wave function of
the harmonic oscillator is Gaussian. This implies that the
non-oscillatory new modes would have strong deviation from Gaussian.

The theory assumes $\Lambda_g = 0$ on the initial spacelike
hypersurface when the universe was born as a classical spacetime.  It
is an interesting question how to treat this for fluctuations
classicalized when they become superhorizon during inflation.
Ultimately the theory of quantum gravity would be required to
answer this question, but it is reasonable to assume
$\delta\Lambda_g = 0$ as the initial condition at classicalization
also for such metric perturbations.  It should be noted that the
initial hypersurface to determine the field equations may also be
perturbed from the background isotropic FLRW metric, which may be
different for different wavelength modes. Then metric perturbations in
various wavelengths cannot be treated in a single synchronous
coordinate system, but the synchronous coordinates in which
$\Xi^{\mu\nu}$ takes the simple form of eqs.
(\ref{eq:Xi_1})--(\ref{eq:Xi_3}) may be different for different
wavelength modes. Systematic formulations to treat such various modes
in a single coordinate system are interesting for future work, but
here evolution of a single wavelength mode is considered for
simplicity.

Consider a mode of metric perturbation in its synchronous coordinate
system.  Because $\delta \Lambda_g = 0$ throughout the spacetime
within the linear theory for the FLRW metric, evolution of $h_{ij}$ is
determined by the $ij$ components of linearized EFE.  Classicalization
occurs when the mode becomes superhorizon, and hence we can ignore
spatial derivative terms in later evolution.  In this case the
equations for the metric perturbations become $\ddot h_{ij} = - 3 H
\dot h_{ij}$, and hence all components of $h_{ij}$ are quickly frozen
and become constant with time. After inflation, such frozen
fluctuations gradually enter the Hubble horizon from shorter
wavelength modes. Then the new modes of metric anisotropy should
affect the observational signal of CMB B-mode polarization with
amplitudes similar to the ordinary two gravitational wave modes, on
scales larger than the horizon at the recombination.  Especially, the
mode $(h_{11} + h_{22})$ cannot be erased by gauge transformations as
discussed in \S\ref{sec:linear_theory}.  The gravitational wave modes
start to oscillate after they become subhorizon, and are damped on the
time scale of $H^{-1}$ by the frictional term in the equations of
motion. Though the new modes are not oscillatory, their evolution
would also change after entering the horizon.  Quantitative evolution
must be calculated by a careful treatment of perturbation equations in
an expanding universe with coupling to matter fluctuations, which is
beyond the scope of this paper.

\section{Concluding remarks}

In this work a new set of gravitational field equations was derived
from the action principle, in which the Lagrangians are the same as
standard general relativity but four constraints are imposed on
$\delta g_{\mu\nu}$ to extract six DOFs of gravity.  Then the four
constraints on initial conditions in EFE are removed.  This is
motivated by a principle that the gravity theory should be able to
describe spacetime evolution starting from free initial conditions on
the initial spacelike hypersurface. Such field equations were originally
proposed in Paper I, but here a new version was derived by replacing
the three of the four constraints with more physically motivated ones,
requesting that geodesics remain geodesic against variations.  Another
constraint of the unimodular condition is unchanged.  As a result, the
theory becomes nonlocal with integro-differential field equations.
The constraints of keeping geodesics were introduced independently of
the motivation of free initial conditions, but they naturally lead to
the field equations similar to those of Paper I and the principle of
free initial conditions is satisfied.  Furthermore, the instability
problem in the version of Paper I naturally disappears. These results
lend some credence to the proposed theory as a candidate of the true
gravity theory.

The field equations (\ref{eq:new-field-eqs}) are given in a covariant
form, which are derived from the covariant condition of the least
action, but the general principle of relativity is violated because
the tensor $\Xi^{\mu\nu}$ takes a simple form in preferred coordinate
systems defined by the initial spacelike hypersurface when a classical
spacetime is born. At the cost of this, the theory allows free initial
conditions about the metric, their first time derivatives, and matter
distribution. Moreover, the cosmological constant problem is solved in
this theory.  One may think that the theory is similar to the
Einstein-Aether (EA) theory in which a timelike vector field is
introduced and hence there is a preferred reference frame
\cite{Jacobson:2008,Clifton:2011}.  However, in the present theory the
preferred frame is defined by a synchronous coordinate system with
respect to the initial hypersurface, and $u^\mu$ is not a physical
dynamical field.  The vector field $\xi^\mu$ is Lagrange multipliers
rather than a dynamical field including kinetic terms, unlike the
vector field in the EA theory.  The concept of the proposed theory is to
increase DOFs of metric dynamics by removing the four constraints on
initial conditions in EFE, rather than introducing new physical
fields.  The proposed theory does not include any new adjustable
parameters, while the EA theory includes several.

The theory is indistinguishable from EFE+$\Lambda$ after inflation as
a classical theory, and hence this theory passes all
observational/experimental tests supporting general relativity and the
standard cosmological model with $\Lambda$ and cold dark matter.
However, it has a particular prediction about the primordial metric
anisotropy generated by quantum fluctuations during inflation. Because
of the increased DOFs of metric dynamics, new modes of fluctuation
would be generated including $h_{11} + h_{22}$ in plane wave solutions
to the $x^3$ direction, which is prohibited in standard general
relativity.  The two gravitational wave modes predicted by EFE are
unchanged, and amplitudes of the new modes are similar to
those. However, the new modes are non-oscillatory and highly
non-Gaussian, unlike the ordinary gravitational waves. These
predictions may be tested by future observations seeking for CMB
B-mode polarization. A fundamental assumption here is that
gravitational fields should be quantized in the same way as other
fields, which may not be true. Another possibility is then that
primordial metric anisotropy (including ordinary gravitational waves)
is not generated at all by quantum metric dynamics during inflation.
The solution to the cosmological constant problem by this theory is
still valid even in such a case, though the theory would become
indistinguishable from EFE$+\Lambda$ within the present
Hubble horizon.





\end{document}